\documentclass[final,3p]{elsarticle}

\usepackage{amssymb}

\journal{New Astronomy Reviews}

\begin{document}

\begin{frontmatter}

\title{The Physical Properties of Red Supergiants}

\author{Emily M. Levesque\footnotemark }

\address{Institute for Astronomy, University of Hawaii, 2680 Woodlawn Dr., Honolulu, HI 96822; emsque@ifa.hawaii.edu}

\begin{abstract}
Red supergiants (RSGs) are an evolved He-burning phase in the lifetimes of moderately high mass ($10-25M_{\odot}$) stars. The physical properties of these stars mark them as an important and extreme stage of massive stellar evolution, but determining these properties has been a struggle for many years. The cool extended atmospheres of RSGs place them in an extreme position on the Hertzsprung-Russell diagram and present a significant challenge to the conventional assumptions of stellar atmosphere models. The dusty circumstellar environments of these stars can potentially complicate the determination of their physical properties, and unusual RSGs in the Milky Way and neighboring galaxies present a suite of enigmatic properties and behaviors that strain, and sometimes even defy, the predictions of stellar evolutionary theory. However, in recent years our understanding of RSGs, including the models and methods applied to our observations and interpretations of these stars, has changed and grown dramatically. This review looks back at some of the latest work that has progressed our understanding of RSGs, and considers the many new questions posed by our ever-evolving picture of these cool massive stars.
\end{abstract}


\end{frontmatter}

\section{Introduction}
Red supergiants (RSGs) \footnotetext{Predoctoral Fellow, Smithsonian Astrophysical Observatory, 60 Garden St., Cambridge, MA 02139} are a He-burning evolutionary phase in the lifetimes of moderately massive ($10M_{\odot} \lesssim M \lesssim 25M_{\odot}$) stars. According to the Conti (1976) scenario for the evolution of massive stars, and its subsequent illustration in Massey (2003), RSGs are the end result of a nearly horizontal evolution across the Hertzsprung-Russell (H-R) diagram as their blue H-burning predecessors leave the main sequence and cross the ``yellow void", passing through the very short-lived yellow supergiant stage. In some cases this is the terminal stage for massive stars, which spend a significant fraction of their time as RSGs before ending their lives as hydrogen-rich Type II supernovae. In other more massive cases, stars will spend a portion of their He-burning lifetimes as RSGs but then evolve back across the H-R diagram, passing once again through the brief yellow supergiant phase and exploding as either blue supergiants or Wolf-Rayet (W-R) stars depending on their initial masses and mass loss rates.

The assumed physical properties of RSGs mark them as a unique and extreme phase of massive stellar evolution. They have the largest physical size of any stars, and their very cool effective temperatures ($T_{\rm eff}$) and extended atmospheres lead to a spectrum that is dominated by molecular absorption lines. The latter two of these characteristics both pose a challenge to the development of accurate stellar atmosphere models; the extended RSG atmospheres, with their large scale heights, invalidate the typical assumption of a plane parallel atmospheric geometry, and their cool temperatures demand a careful and complete treatment of molecular opacities. The temperatures of RSGs also result in significant negative and temperature-sensitive bolometric corrections on the order of a few magnitudes. As a result, determining these stars' luminosities is dependent on careful $T_{\rm eff}$ measurements, making an accurate picture of their physical properties vital to any attempts at placing these stars in their proper place on the H-R diagram.

Such measurements are a challenge in and of themselves; the physical properties of RSGs have remained poorly understood, and at odds with the predictions of stellar evolutionary theory, until recently. This is due in part to the relatively sparse number of nearby RSGs in the Milky Way, difficulties in differentiating RSGs from red foreground stars when studying extragalactic samples, and significant complexities introduced as a result of these stars' mass loss rates and dusty circumstellar environments. Their extreme physical properties also make them very difficult to model in detail. Despite their importance in massive stellar evolution, RSGs were generally ignored for many years by the massive star community, with a few important exceptions (e.g., Humphreys 1978, 1979a, 1979b, Humphreys \& McElroy 1984, Elias et al.\ 1985). Further investigations of RSGs in recent years have made important strides towards answering many outstanding questions about these stars. These same studies have also introduced a number of new questions and revealed the true complexity of RSGs and their critical place in the grand scheme of massive stellar evolution.

In this review I look back at the latest advances in our understanding of RSG physical properties, beginning with the methods that must be used in order to photometrically and spectroscopically identify RSGs as massive stars in the Milky Way or other nearby galaxies (Section 2). Recent work has determined RSG effective temperature scales, bolometric luminosities, and their resulting placement on the Hertzsprung-Russell diagram, revealing interesting metallicity effects on the physical properties of these stars and posing new questions about their masses, luminosities, and lifetimes (Section 3). We have gained a new and more detailed understanding of the mass loss rates and mechanisms of these stars, as well as new insights into the dust produced by RSGs (Section 4). Finally, studies of unusual RSGs in the Milky Way and the Magellanic Clouds have highlighted unusual behaviors and properties of massive stars that challenge our current picture of stellar evolution (Section 5). Finally, I consider the questions that are currently being posed as a result of our improved understanding of RSG physical properties and their importance in stellar evolution (Section 6).

\section{Identifying Red Supergiants}
Massive stars with $10M_{\odot} \lesssim M \lesssim 25M_{\odot}$ spend a significant fraction of their He-burning lifetimes as RSGs. The evolutionary pathway of RSGs combined with the interior stellar physics unique to massive stars mark an important distinction between $\gtrsim 10M_{\odot}$ RSGs and $\lesssim 10M_{\odot}$ AGB stars, another population of luminous red stars evolved from less massive predecessors. However, distinguishing between RSGs and AGBs on the H-R diagram is difficult. There is some overlap between low-luminosity RSGs and the general AGB population (Brunish et al.\ 1986), which can be addressed in part by imposing a luminosity cut-off on stars that can be considered RSGs (e.g. Massey \& Olsen 2003). However, the most luminous AGB stars, sometimes referred to as ``super"-AGBs (e.g. Siess 2006, 2007; Eldridge et al.\ 2007; Poelarends et al.\ 2008) could potentially occupy the same region of the H-R diagram as typical RSGs. The potential for such confusion is due to the degeneracy between mass and luminosity near the evolutionary limits of the Hayashi track, the rightmost point on the H-R diagram where a star can remain in hydrostatic equilibrium (Hayashi \& Hoshi 1961). As a result, AGB stars could potentially contaminate color- and luminosity-selected samples of RSGs.

Identifying extra-galactic RSG populations also presents a significant challenge, due to the hazards of contamination from red foreground stars. This difficulty is noted by Humphreys \& Sandage (1980), who conducted a detailed photometric survey of M33 and identified the brightest red stars in the sample. The distribution of these stars as compared to the brightest blue were not the same, a finding at odds with the expectations that both red {\it and} blue massive stars in M33 would be clustered together in the same OB associations. Humphreys \& Sandage (1980) acknowledged that contamination by foreground dwarfs could be a possible contributor to this phenomenon. This phenomenon could not be explained by differences in the ages of blue and red supergiants, as this would amount to no more than a few million years - at a drift speed of 30 km s$^{-1}$ this would amount to a drift of only $\sim$1.5 arcminutes in M33, not enough to explain the apparent disagreement.

The discrepancy was eventually resolved by Massey (1998), which found that $\sim$50\% of the red stars included in the Humphreys \& Sandage (1980) sample were in fact foreground red dwarfs. Massey (1998) established a means of discriminating between the low-gravity background RSGs and high-gravity foreground dwarf contaminants. Placing the full sample of red stars on a ($B-V$) vs ($V-R$) color-color diagram reveals a clear separation in ($B-V$) between the M33 RSGs and the foreground Milky Way red dwarfs. This is a consequence of enhanced line blanketing effects at lower surface gravities, which are particularly influential in the $B$ band as a result of the number of weak metal lines in that wavelength regime.

It is important to note that the inclusion of red giants in the halo of the Milky Way could also potentially contaminate samples of extragalactic RSGs. Levesque et al.\ (2007) carefully consider this issue for the case of halo red giants in the direction of the Large Magellanic Cloud (LMC). They find the likelihood of halo giant contamination in the direction of the LMC to be less than 1\% and confirm the membership of their RSG sample based on the kinematic analysis of the LMC by Olsen \& Massey (2007). Massey et al.\ (2009) again address this issue for the RSG population of M31, performing a careful analysis of radial velocities for the RSG candidate spectra in their sample and confirming that all stars in the sample have velocities consistent with their locations in M31, as determined from Rubin \& Ford (1970). Both of these examinations highlight the low probability of halo contamination in extragalactic RSG samples; however, they also emphasize that detailed kinematic analyses of the presumed host galaxies are required to rigorously address this issue.

\section{Red Supergiants and the H-R Diagram}
\subsection{Galactic Red Supergiants}
Massey (2003) and Massey \& Olsen (2003) noted that RSGs appeared to be at odds with the current predictions of stellar evolutionary theory. The Galactic-metallicity evolutionary tracks of the Geneva group (Schaller et al.\ 1992, Meynet et al.\ 1994) failed to extend to temperatures cool enough to accommodate the Galactic RSG samples of Humphreys (1978) and Humphreys \& McElroy (1984) on the Hertzsprung-Russell (HR) diagram. A similar discrepancy was found in the lower-metallicity Magellanic Clouds; the evolutionary tracks of Schaerer et al.\ (1993) (z=0.008) and Charbonnel et al.\ (1993) (z=0.004) did not agree with RSG samples from Elias et al.\ (1985) and Massey \& Olsen (2003).

Such a disagreement was not surprising, considering the many challenges that RSGs present to the evolutionary models. Initially, the problem was attributed to difficulties in accurately modeling mixing-length. The velocities of convective layers in RSGs are nearly sonic, and even supersonic in the atmospheric layers, which produces shocks (Freytag et al.\ 2002) and invalidates mixing-length assumptions. This also results in an asymmetric photosphere and a poorly defined radius, a phenomenon demonstrated in high angular resolution optical and near-infrared observations of Betelgeuse (e.g., Young et al.\ 2000, Tatebe et al.\ 2007, Ohnaka et al.\ 2009, Kervella et al.\ 2009). In addition to asymmetries in these stars' atmospheres, the highly extended atmospheres are at odds with the plane-parallel geometry assumptions of stellar atmosphere models. Finally, the cool effective temperatures of RSGs demand models that include accurate opacities for molecular transitions, such as the TiO bands that dominate their spectra.

Levesque et al.\ (2005) considered the additional fact that the derived physical properties of RSGs, were also highly uncertain. The position of an RSG on the H-R diagram is primarily dictated by the star's $T_{\rm eff}$. These cool stars have significant $T_{\rm eff}$-dependent bolometric corrections; as a result, a 10\% error in $T_{\rm eff}$ corresponds to a factor of 2 error in their bolometric luminosities ($M_{\rm bol}$; Kurucz 1992, Massey \& Olsen 2003), making accurate determinations of $T_{\rm eff}$ critical to proper placement on the H-R diagram.

Unfortunately, careful determinations of $T_{\rm eff}$ scales for RSGs have been difficult to derive. A lack of nearby RSGs precludes the use of measured stellar diameters to generate a basic relation between $T_{\rm eff}$ and spectral subtype, as has been done in the past for red giants. Humphreys \& McElroy (1984) produced a $T_{\rm eff}$ scale for Galactic RSGs; $T_{\rm eff}$ was determined by assuming a blackbody continuum and using broadband colors to assign $T_{\rm eff}$ based on the small sample of nearby RSGs with measured diameters (Johnson 1964, 1966, Lee 1970). However, the gravity-dependent line blanketing effects described by Massey (1998) strongly affect the ($B-V$) colors of these stars. In the case of RSGs in the Magellanic Clouds, Massey \& Olsen (2003) shifted the Dyck et al.\ (1996) scale for red {\it giants} (determined from interferometric data and lunar occultation measurements) down by 400 K based on the more limited Dyck et al.\ (1996) RSG data. They note that this is a very uncertain determination that does not take into account, for example, potential metallicity effects, and stress that a more careful scale derived using accurate spectrophotometry and RSG-appropriate surface gravities is needed. 

The new generation of the MARCS stellar atmosphere models (Gustafsson et al.\ 1975, Plez et al.\ 1992)  include improved treatments of the effects of opacities of oxygen-rich molecules, especially TiO (Plez 2003, Gustafsson et al.\ 2003, Gustafsson et al.\ 2008). In conjunction with the spectrophotometric observations proposed in Massey \& Olsen (2003), these models could be used to make robust determinations of $T_{\rm eff}$ based on the rich TiO bands that dominate RSG spectra, particularly M supergiants. This in turn made the models ideal tools for constructing a new $T_{\rm eff}$ scale for RSGs.

Levesque et al.\ (2005) obtained moderate-resolution spectrophotometry of 74 Galactic RSGs, and used the MARCS stellar atmosphere models to determine a new $T_{\rm eff}$ scale for these stars, along with measurements of $A_V$ and surface gravity. Fitting of the models to determine $T_{\rm eff}$ was based primarily on the strength of the same rich temperature-sensitive TiO bands that are used to assign RSG spectral subtypes ($\lambda\lambda$ 6158, 6658, 7054; see Jaschek \& Jaschek 1990), with TiO bands further in the blue ($\lambda\lambda$ 5167, 5448, 5847) serving as secondary confirmations of the quality of the fit. The $T_{\rm eff}$ determined from model fitting was also used to derive $M_{\rm bol}$ for the RSG sample, using the $T_{\rm eff}$-dependent bolometric corrections in the $V$ band derived from the MARCS stellar atmosphere models.

Levesque et al.\ (2005) also employed an alternative means of determining $T_{\rm eff}$ from RSG ($V-K$)$_0$ colors. Josselin et al.\ (2000) show that RSG $K$ magnitudes are useful in deriving $M_{\rm bol}$. The $K$-band bolometric correction is relatively constant with respect to $T_{\rm eff}$ and surface gravity, $K$ magnitudes are less sensitive to reddening effects, and RSGs are much less variable in the $K$ band than the $V$ band (making $K$ photometry the best means of determining $M_{\rm bol}$ for these stars). By determining the relationship between $T_{\rm eff}$, ($V-K$)$_0$, and the bolometric correction at $K$ based on the MARCS models, Levesque et al.\ (2005) determined alternative values for $T_{\rm eff}$. In the end, $T_{\rm eff}$ determinations from the molecular band strengths and the ($V-K$)$_0$ colors agreed to within 100 K, with the ($V-K$)$_0$ colors generally giving slightly higher effective temperatures than those derived from spectral fitting.

The resulting new $T_{\rm eff}$ scale brought the Milky Way RSG population into excellent agreement with the predictions of the new Geneva group evolutionary models, which included treatments of stellar rotation (see Figure 1). This work also included determination of these stars' stellar radii for the first time. Several RSGs - KW Sgr, Case 75, KY CYg, and $\mu$ Cep - were found to have radii of $\sim1500R_{\odot}$, making these the largest single stars known and coinciding precisely with the predictions of current evolutionary theory for the maximum radius attainable by Galactic RSGs.

\begin{figure}
\includegraphics[width=8cm]{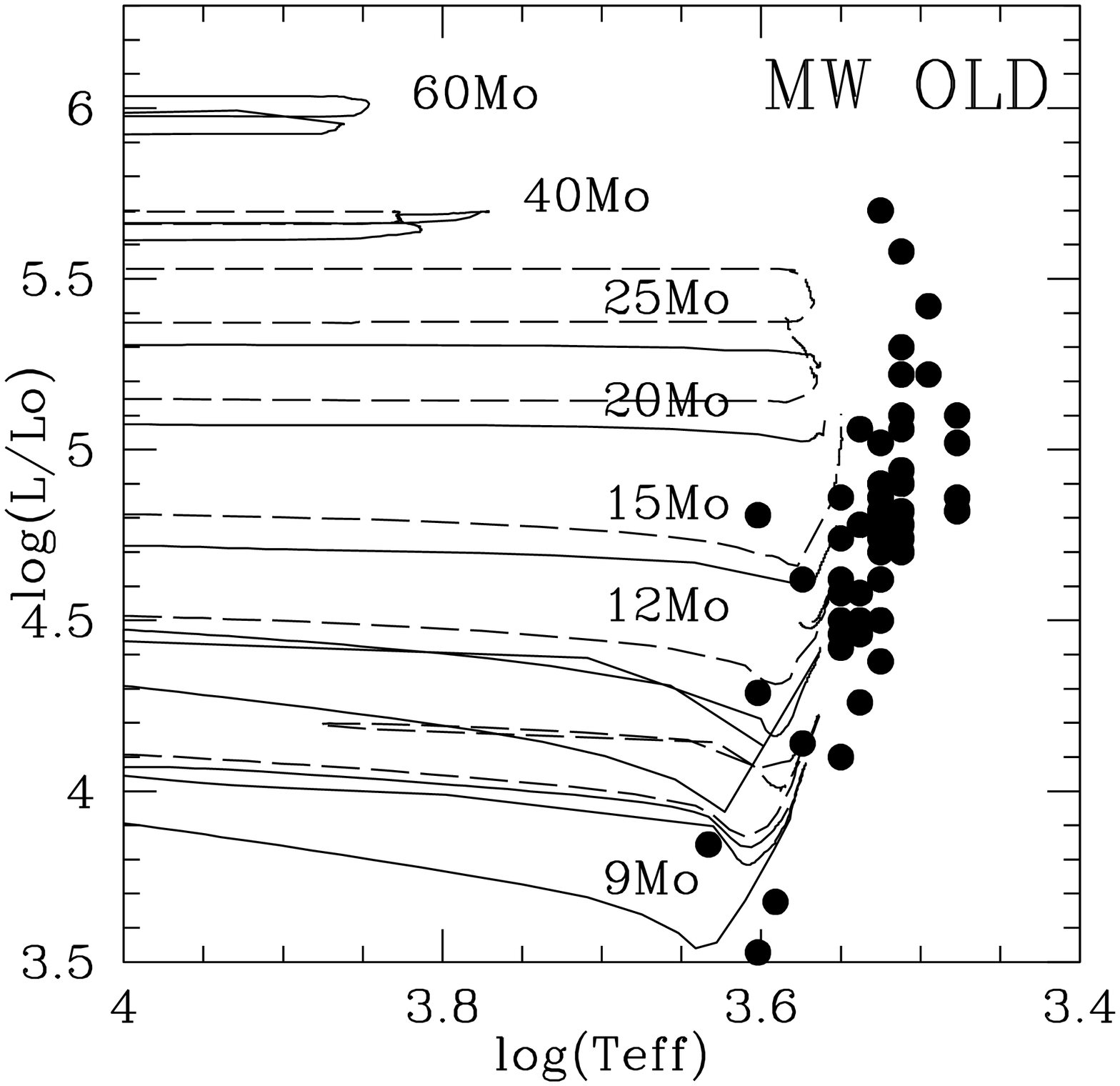}
\includegraphics[width=8cm]{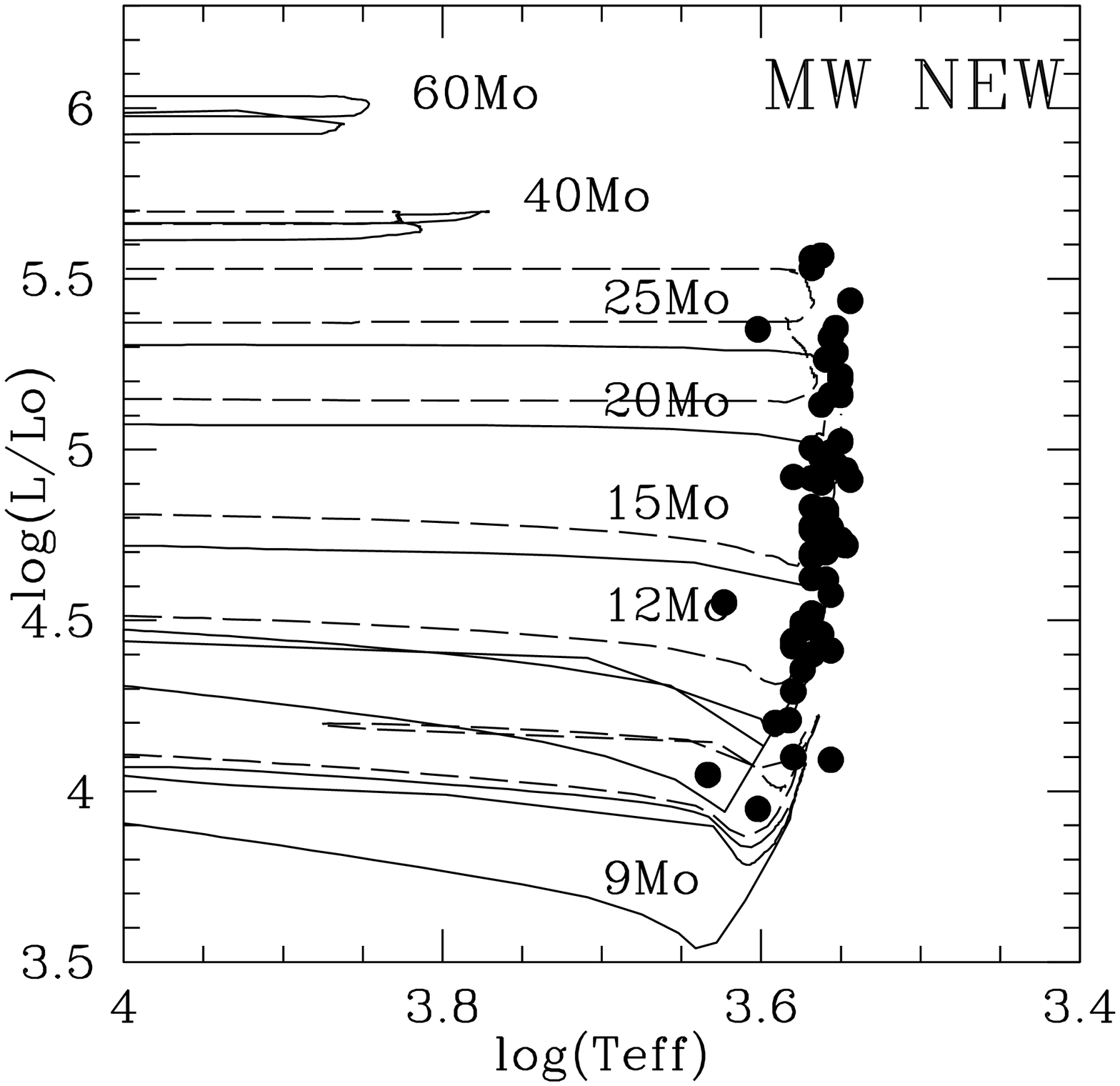}
\caption{Comparison of Galactic RSGs with the evolutionary tracks, adapted from Levesque et al.\ 2005. The H-R diagram for the Milky Way compares the predictions of the evolutionary tracks to the position of RSGs from Humphreys (1978), adopting the effective temperatures and bolometric corrections of Humphreys \& McElroy (1984) ({\it left}), and to RSGs from Levesque et al.\ (2005) ({\it right}). The evolutionary tracks are from Meynet \& Maeder (2003) and include both non-rotating tracks (solid lines), and tracks that assume an initial rotation velocity of 300 km s$^{-1}$ (dashed lines). The Galactic RSGs from Levesque et al.\ (2005) show greatly improved agreement with the tracks.}
\end{figure}

\subsection{Metallicity Effects on the H-R Diagram}
\subsubsection{Effective Temperatures}
Following the redetermination of a Galactic $T_{\rm eff}$ scale for RSGs, Levesque et al.\ (2006) performed a similar analysis on spectrophotometry of 36 RSGs in the LMC and 37 RSGs in the Small Magellanic Cloud (SMC). The $T_{\rm eff}$ scales for the Magellanic Clouds produced comparable results to the analysis in the Milky Way, resolving the disagreement between RSGs and evolutionary tracks noted in Massey \& Olsen (2003). While the LMC RSGs were brought into excellent agreement with the Geneva evolutionary tracks (Schaerer et al.\ 1993, Meynet \& Maeder 2005), there was an improved but not wholly satisfactory agreement with the SMC tracks (Charbonnel et al.\ 1993, Maeder \& Meynet 2001). The SMC sample showed a considerably larger spread in $T_{\rm eff}$ across a given luminosity as compared to their LMC and Milky Way counterparts. However, such a spread is not entirely surprising due to the expected enhancement of rotational mixing effects in stars at these lower metallicities (Maeder \& Meynet 2001).

The metallicity effects on the RSG population, and the challenges they pose to stellar evolutionary theory, are not limited to disagreements between $T_{\rm eff}$ and the predictions of the evolutionary tracks. When comparing RSGs in the Milky Way and the Clouds, Elias et al.\ (1985) noted an interesting shift in the spectral types of these stars, with the average RSG spectral subtype shifting toward earlier types at lower metallicities. More precisely, the average RSG subtype is found to be K5-K7 in the $Z = 0.2Z_{\odot}$ SMC, M1 I in the $Z = 0.5Z_{\odot}$ LMC, and M2 I in the $Z = Z_{\odot}$ Milky Way (Massey \& Olsen 2003). Levesque et al.\ (2006) present two distinct explanations for this shift. Firstly, the TiO bands that dictate the spectral type of M and late-K RSGs are sensitive to chemical abundance as well as temperature. A RSG with a $T_{\rm eff}$ of 3650 K, for example, would be assigned a spectral type of M2 I in the Milky Way, M1.5 I in the LMC, and K5-M0 I in the SMC, based purely on TiO bands that become weaker at lower metallicity (Figure 2a). Secondly, the Hayashi limit imposes a restriction on how cool, and hence how late-type, RSGs are permitted to be while remaining in hydrostatic equilibrium. This limit shifts to warmer temperatures, and therefore earlier spectral types, at lower metallicity; a 15-25M$_{\odot}$ RSG at the coolest point of its evolution will be about 100-150 K warmer in the LMC as compared to the the Milky Way, and about 500 K warmer than the Milky Way in the SMC (Figure 2b). This 500 K difference is actually in excess of the 350 K change seen in observations of SMC RSGs.
\begin{figure}
\includegraphics[width=8cm]{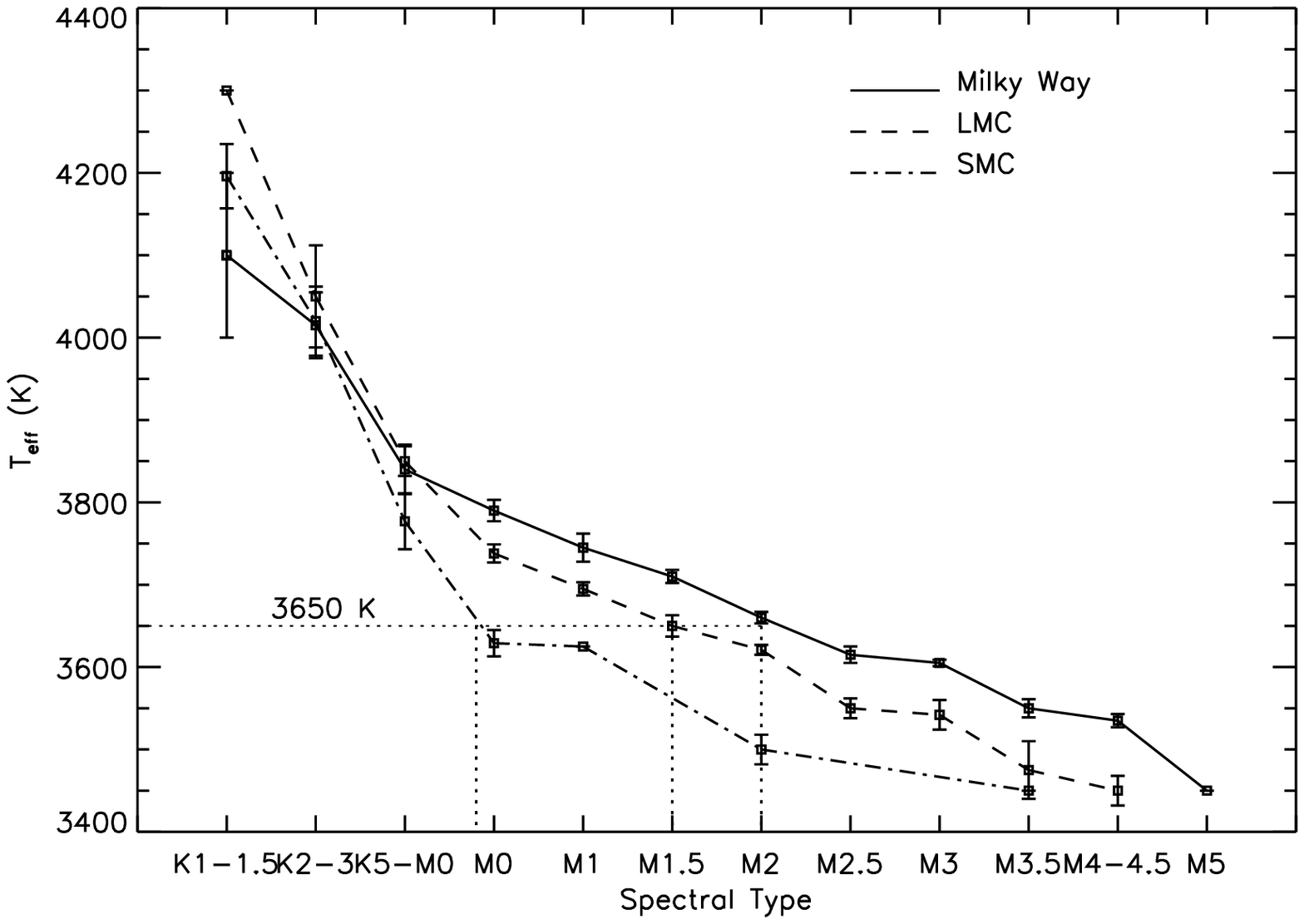}
\includegraphics[width=8cm]{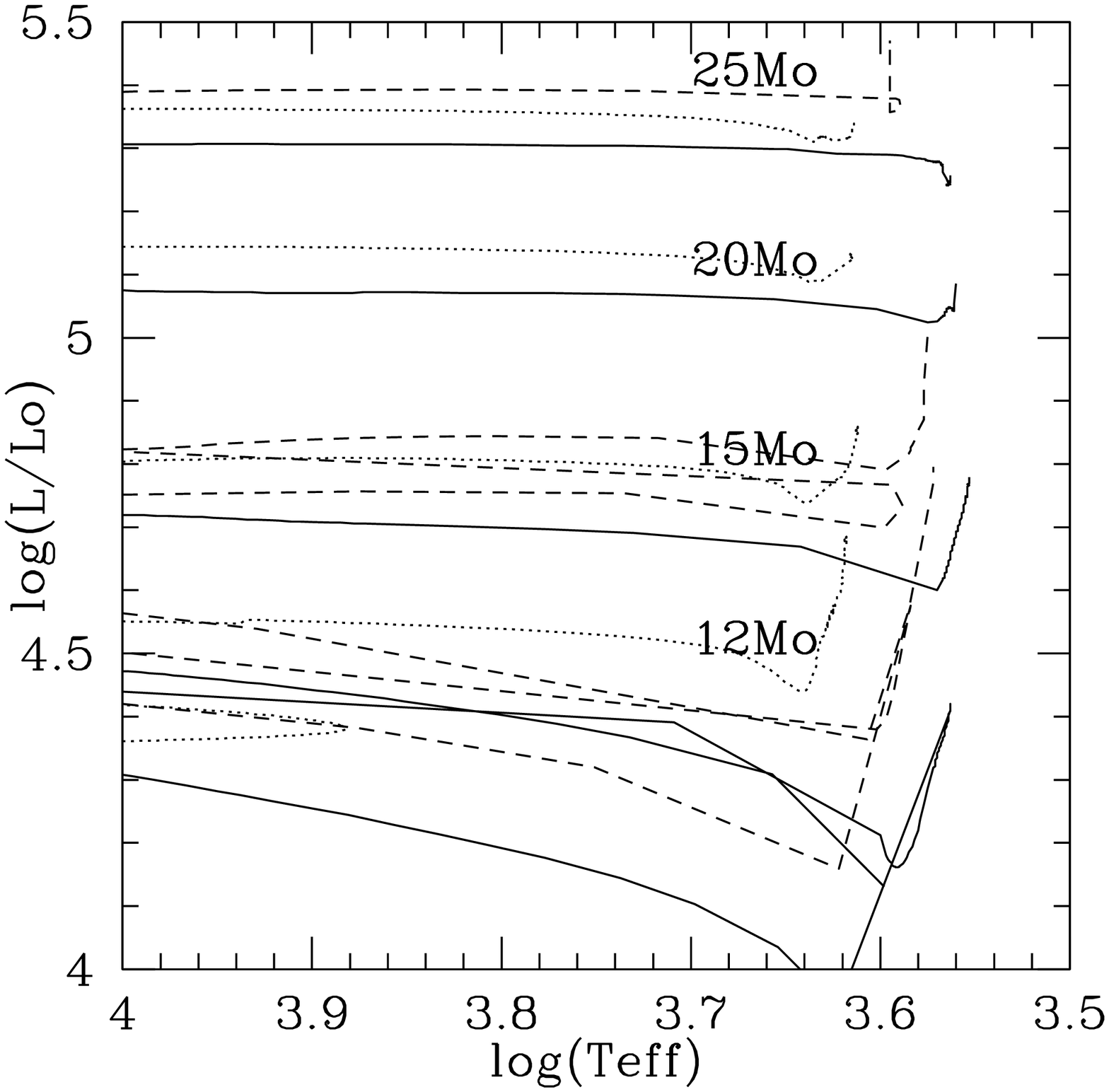}
\caption{Explaining the shift in average RSG spectral types with metallicity, adapted from Levesque et al.\ 2006. {\it Left}: The $T_{\rm eff}$ scales for RSGs in the Milky Way (solid line), LMC (dashed line), and SMC (dashed-dotted line). A dotted line of constant $T_{\rm eff}$ is drawn at 3650 K, and illustrates that the same $T_{\rm eff}$ corresponds to a spectral type of M2 in the Milky Way, M1.5 in the LMC, and K5-M0 in the SMC, a result of abundance effects on the strengths of the TiO lines. {\it Right}: A comparison of non-rotating evolutionary tracks for the Milky Way (solid line), LMC (dashed line), and SMC (dotted line), illustrating the shift in the Hayashi limit to warmer temperatures at lower metallicities. The evolutionary tracks are from Meynet \& Maeder (2003) for the Milky Way, Schaerer et al.\ (1993) for the LMC, and Charbonnel et al.\ (1993) for the SMC.}
\end{figure}

Most recently, Massey et al.\ (2009) fit the MARCS stellar atmosphere models to spectrophotometry of 16 RSGs in M31, a galaxy whose metallicity is currently under debate (see Crockett et al.\ 2006). This small sample of RSGs agrees with both solar and 2$\times$ solar metallicity tracks; however, Massey et al.\ (2009) emphasize that a larger survey of M31 RSGs could be helpful in settling the debate surrounding this galaxy's metallicity. With a larger sample the average spectral subtype in M31 could be determined, a property which could prove quite informative in light of the observed relationship between RSG spectral subtype and host metallicity.

\subsubsection{Lifetimes}
Another challenge metallicity poses to the modeling of RSGs concerns the ratios of blue to red supergiants (B/R) and RSGs to Wolf-Rayet stars (RSG/W-R). Van den Bergh (1968, 1973) first noted that the relative number of blue supergiants and RSGs in nearby galaxies decreased with the galaxies' absolute magnitude. They proposed that this was potentially due to a corresponding decrease in chemical abundance, following the direct relation between luminosity and metallicity for galaxies. Humphreys \& Davidson (1979) notes that the B/R in a galaxy or cluster may be indicative of the time spent in different evolutionary stages; this explanation was further expanded upon by Maeder et al.\ (1980), who suggested that the change in B/R is due to lower mass loss rates in low-metallicity environments, which would in turn lead to longer RSG lifetimes.

Maeder et al.\ (1980) also propose that the RSG/W-R ratio should decrease with increasing metallicity, again as a function of abundance-dependent mass loss rates and the corresponding effects on RSG and W-R lifetimes. Massey (2002) confirm this trend, finding a factor of 160 difference in the RSG/W-R ratio over a spread of $\sim$0.9 dex in metallicity (from the SMC to M31, taking M31's oxygen abundances to be log(O/H) + 12 = 9.0 from Zaritsky et al.\ 1994). Evolutionary models for massive stars have not yet fully reproduced the change in these ratios, but models that include treatments of enhanced mass loss (Meynet et al.\ 1994) and stellar rotation (Maeder \& Meynet 2001) have made significant strides in accommodating the observed B/R and RSG/WR ratios in recent years. It is also important to note that B/R and RSG/W-R ratios run the risk of being contaminated by samples such as luminous AGB stars (see Section 2) or under-sampled due to the non-inclusion of warmer (and hence ``yellower") RSGs in low-metallicity environments.

\subsubsection{Maximum Luminosities}
Massey (1998) argue that there is a relation between the maximum luminosities ($L_{\rm max}$) of RSGs and metallicity, with lower-metallicity RSGs having higher luminosities. Like the B/R and RSG/W-R ratios, the dependence of $L_{\rm max}$ on metallicity can be traced back to abundance-dependent mass loss effects. In higher-metallicity environments, it is expected that a massive star of a particular mass will immediately become a Wolf-Rayet star upon leaving the main sequence, a consequence of the high rate of mass loss enabling the outer H and He layers to be shed at a faster rate. By contrast, at lower metallicities a star with the {\t same} mass will evolve through an intermediate, and perhaps even terminal, RSG phase because the outer layers are shed via a much slower mass loss rate. 

Massey et al.\ (2009) derive $M_{\rm bol}$ for RSGs in the Milky Way, the Magellanic Clouds, and M31, and find that the most luminous RSGs have consistent log $L/L_{\odot} \sim$ 5.2-5.3 across the $\sim$0.9 dex metallicity spread between M31 and the SMC. This is at odds with the expectation that $L_{\rm max}$ should vary across these metallicities. However, these results should not necessarily be seen as a challenge to evolutionary theory. The shift in the Hayashi track with metallicity complicates any analyses that require a complete picture of the RSG population in galaxies beyond the Milky Way. In low-metallicity galaxies such as the SMC, for example, massive stars are not expected to evolve past the K or early-M spectral type, a result of the higher-$T_{\rm eff}$ limitations of the Hayashi track restricting their effective temperatures to $>$4500 K for rotating evolutionary models (Meynet \& Maeder 2005) or an even warmer limit of $>$5600 K for non-rotating models (Charbonnel et al.\ 1993). As a result, an accurate sample of the late-type massive star population at these metallicities must therefore include {\it yellow} supergiants as well as K- and M-type RSGs. Massey (2002) notes that including both M-type {\it and} K-type supergiant candidates when determining B/R for the LMC and SMC significantly alters the ratios, and Massey et al.\ (2009) point out that yellow supergiants must be included in samples of low-metallicity late-type stars to ensure a proper determination of $L_{\rm max}$. Recent strides have been made in identifying extragalactic yellow supergiant populations, although such surveys are challenging due to the very short lifetimes of the yellow supergiant phase and the dominance of foreground contamination, expected to be between 50\% and 95\% for galaxies in the local group (Massey et al.\ 2006a, 2007a, Drout et al.\ 2009). 

Finally, in a comparison with the evolutionary tracks, the $L_{\rm max}$ determined by Massey et al.\ (2009) correspond a maximum RSG mass of $\sim$25-30$M_{\odot}$. Smartt et al.\ (2009) note an apparent discrepancy between this implied maximum mass and the maximum RSG masses observed as Type II-P supernova progenitors. While observations of $L_{\rm max}$ imply a maximum stellar mass of $\sim$25-30$M_{\odot}$ for RSGs based on comparisons with the evolutionary tracks, only RSGs with apparent initial masses of $\lesssim17M_{\odot}$ are detected as Type II-P progenitors. Smartt et al.\ (2009) refer to this as the ``red supergiant problem", noting that the explosive deaths of stars with masses between 17$M_{\odot}$ and 25-30$M_{\odot}$ are not observed. However, it is very important to note that it is very difficult to draw robust conclusions about RSGs mass ranges based solely on measurements of $L_{\rm max}$. Mass and luminosity are degenerate around the Hayashi track region of the H-R diagram, as the evolutionary tracks become nearly vertical and the luminosities of stars at different masses begin to overlap considerably. This makes detailed determinations and analyses of RSG masses difficult.

\section{Dust Production in Red Supergiants}
The first detections of circumstellar dust shells around RSGs came in the late 1960s. Johnson (1968) proposed the presence of an extensive circumstellar cloud surrounding the extreme RSG NML Cyg, based on infrared spectroscopy of the star showing a large infrared excess and a very high luminosity. Hyland et al.\ (1969) also predicted the presence of a circumstellar dust shell around the extreme RSG VY CMa based on infrared photometry and spectra. This work also found that the extinction curve for the circumstellar dust surrounding VY CMa differed from that of normal interstellar dust, showing larger relative extinction in the infrared as compared to the optical and consistent with a larger-than-average grain size. Snow et al.\ (1987) found similar results when examining the circumstellar envelope surrounding the RSG binary $\alpha$ Sco, indicating that the dust grains are large and consist primarily of silica. Hagen (1978) did a detailed analysis of circumstellar gas around M giants and supergiants, based on studies of line profiles, and found that mass loss in these stars was {\it not} being driven by radiation pressure on the dust grains, as had been previously believed; see also Hagen et al.\ (1983). Finally, work by Stencel et al.\ (1988, 1989) revealed that circumstellar dust shells are not unique to the most extreme RSGs, but are also common in the Galactic RSG population as a whole.

Circumstellar dust shells are formed as a consequence of grain condensation during stellar mass loss. Massive stars are thought to lose more than half of their mass after they evolve off of the main sequence (e.g., Stothers \& Chin 1996), and much of this mass loss has been found to occur during the RSG phase. Danchi et al.\ (1994) found variations in the distance of the circumstellar dust shells from RSG photospheres, and cited this as evidence of sporadic mass loss episodes in RSGs separated by several decades. Salasnich et al.\ (1999) modeled a new luminosity-dependent mass loss rate for Magellanic Cloud RSGs that was $\sim$2-5 times higher than previous estimates and incorporated a metallicity-dependent component. By contrast, Josselin et al.\ (2000) surprisingly found no clear correlation between luminosity and mass loss rate across a large sample of Galactic RSGs. However, this was determined by adopting distances to these stars based on the individual spectroscopic parallaxes of Humphreys (1978), which leads to a poor approximation of these stars' luminosities.

The Levesque et al.\ (2005) survey of Galactic RSGs noted that many of these stars' spectra had excess flux in the near-ultraviolet (NUV) as compared to the predictions of the MARCS stellar atmosphere models (for an example, see Figure 3). Upon closer examination in Massey et al.\ (2005), this excess flux was found to be closely correlated with the amount of excess reddening present in the RSGs relative to their OB associations. In addition, Massey et al.\ (2005) also revisit the mass loss rate determined by Josselin et al.\ (2000) by taking the RSG distances to be the average cluster values for their OB association. The result produced a mass loss rate for Galactic RSGs that is dependent on luminosity. More recently, Bennett et al.\ (2009) use IR, optical, and ultraviolet spectra to demonstrate that the circumstellar dust surrounding the Galactic RSG $\mu$ Cep does not follow a standard reddening law, instead finding $R_V \gtrsim 6$, implying that this dust does not follow the standard $R_V = 3.1$ Cardelli et al.\ (1989) reddening law found for the diffuse Galactic ISM. 
\begin{figure}
\includegraphics[width=15cm]{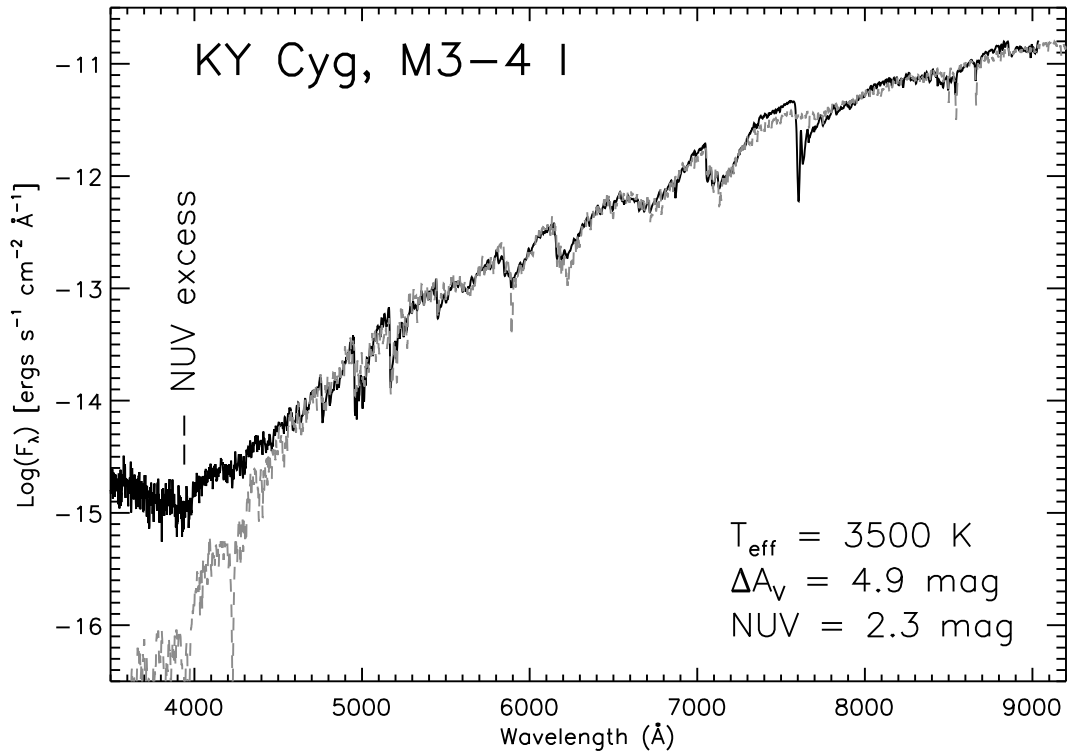}
\caption{The effects of circumstellar dust in the NUV, adapted from Massey et al.\ (2005). The observed spectral energy distribution (SED) of KY Cyg (black) is compared to the best-fit synthetic SED from the MARCS stellar atmosphere models. The fit has been corrected for extinction by ``reddening" the model, rather than dereddening the observed SED of the RSG. KY Cyg shows a considerable amount of excess flux in the NUV as compared to the reddened MARCS model. We can also see that there is a considerable amount of excess reddening present in the direction of KY Cyg, with a difference of 4.9 mag as compared to its host OB association, Cyg OB1.}
\end{figure}

There are several important arguments for improving our current understanding of the dust produced by RSGs. The most immediate concern deals with the effects that this dust can have on the assumed $M_{\rm bol}$ for dust-enshrouded RSGs. These effects are largely dependent on the geometry of the dusty envelope. The Levesque et al.\ (2009a) examination of WOH G64, a dust-enshrouded RSG in the LMC, find that deriving $M_{\rm bol}$ from the star's spectrum leads to a considerable overestimation, as it fails to take into account the $\sim$0.5 mag contribution from that star's dusty torus. A similar difficulty has plagued recent debates surrounding the derived luminosity of VY CMa; for further discussion see Massey et al.\ (2006b), Humphreys et al.\ (2007), and Section 5.1.

The dust production of RSGs is also important in the study of extragalactic ISM environments. In a galaxy such as the Milky Way, or other environments with an underlying old stellar population component, AGB stars and SNe contribute the majority of the dust in the ISM. However, in starburst galaxies at large lookback times where there is no older population of low-mass stars, RSGs are expected to dominate dust production. This should be particularly prevalent in low-metallicity starbursts, where evolved dust-producing Wolf-Rayet stars are rare. Low-metallicity starburst galaxies at high redshifts are currently a source of considerable interest. They are important components in studying the evolution of metallicities and star-formation rates as a function of redshift. They are also precisely the sort of environments that appear to produce long-duration gamma-ray bursts (e.g., Stanek et al.\ 2006, Fruchter et al.\ 2006, Modjaz et al.\ 2008, Kocevski et al.\ 2009, Levesque et al.\ 2009b). As a result, there is great interest in producing proper and detailed photoionization models of these low-metallicity starbursts. This will require accurate treatments of extinction and other dust effects, and demands that the mass loss rates, dust properties, and circumstellar environments of RSGs be examined in more detail.

\section{Unusual Red Supergiants}
In any population of stars, the unusual members and apparent outliers are particularly intriguing and often turn out to be instrumental in uncovering new stellar phenomena. The unusual RSGs VY CMa and NML Cyg were instrumental in early studies of circumstellar dust shells around RSGs, which in turn led to our current interest in these stars' mass loss rates and dust production. As studies of RSGs continue, more and more stars are being uncovered whose physical parameters and dust properties set them apart as unique when compared to the general population.

Studies of RSGs in known binary systems have also made vital contributions to our current picture of these stars' atmospheric properties and evolution. The spectroscopic $\zeta$ Aur RSG binaries can be used to examine mass loss rates (e.g., Che et al.\ 1983, Harper et al.\ 2005), and Snow et al.\ (1987) were able to probe the circumstellar environment and dust properties of M-type RSG binary $\alpha$ Sco in detail. Studies of VV Cep, an eclipsing binary composed of an M-type RSG and a hot B-type companion, have also revealed a great deal about RSG mass loss rates and atmosphere structure (e.g., Bauer et al.\ 1991, 2008). However, for the purposes of this discussion we restrict ourselves to the unusual properties and evolutionary implications of single stars.

\subsection{Dust-Enshrouded Red Supergiants}
A subset of RSGs in the Milky Way (and, more recently, in the Magellanic Clouds, e.g. van Loon et al.\ 1998a, 2005), are set apart from the general RSG population by a unique suite of physical properties that can be attributed to their dust production and luminosity-dependent mass loss. These stars are characterized by thick asymmetric circumstellar nebulae, their designation as extremely bright IR sources, and  OH, SiO, and H$_2$O maser emission. Such RSGs are sometimes referred to as OH/IR sources, due to their strong OH maser activity and high IR luminosities. It should be noted that maser activity in late-type RSGs may not be as unusual as previously thought - in a survey of RSGs in the young Galactic cluster RSGC1 (Figer et al.\ 2006), Davies et al.\ (2008) find evidence of maser activity in the most luminous RSGs in the sample, suggesting that maser emission is a phase of stellar evolution that ignites or intensifies in the latest RSG evolutionary phases, when the stars' mass loss rates are expected to be at their highest. However, the presence of the thick asymmetric dust nebulae surrounding these stars definitively set them apart as distinct from normal RSGs.

These dust nebulae can significantly complicate attempts to determine the central RSG physical properties. A thick circumstellar envelope can lead to the apparent ``veiling" of the central star's spectrum to an observer; the absorption lines in the stellar spectra can be diluted or even fully obscured due to multiple scatterings of the photons in the expanding dust shell (Romanik \& Leung 1981); one extreme example of this appears to be the M31 RSG J004047.84+405602.6 (Massey et al.\ 2009). This phenomenon could easily lead to incorrect estimates of a veiled RSG's spectral type or $T_{\rm eff}$. Even more troublingly, a failure to properly account for contributions from the dust envelope can lead to potential under- or over-estimates of an RSG's luminosity, which, in turn, changes its position on the H-R diagram and its assumed mass loss rate. Two good examples of RSGs whose assumed physical properties have been strongly affected by their circumstellar environments - VY CMa in the Milky Way and WOH G64 in the LMC - are discussed in detail below. In the past several years, observations of both of these stars have led to important new insights into how to properly interpret observations of dust-enshrouded RSGs. Several other Galactic OH/IR supergiants - VX Sgr, S Per, and NML Cyg - are also considered in the context of these recent improvements.

\subsubsection{VY Canis Majoris}
VY CMa has long been recognized as a remarkable and unusual RSG. Its circumstellar nebula was first observed in 1917 (Perrine 1923). For many years this was thought to be a multiple system (See 1897), but later observations revealed that the objects originally thought to be stellar companions were in fact structure within the circumstellar nebula (Herbig 1972, Worley 1972). The nebula itself has been studied extensively since then, and found to be asymmetric and highly structured (e.g., Monnier et al.\ 1999, Smith et al.\ 2001, Smith 2004, Humphreys et al.\ 2005, 2007, Jones et al.\ 2007). The heating of this nebula by the star also makes it one of the brightest 5-20$\mu$m objects in the sky (Herbig 1970a). The star has shown photometric variability on the order of $\pm$2 mag extending back over 200 years (Robinson 1970, 1971). VY CMa also shows the strong H$_2$O, SiO, and OH maser emission that is typical of late-type mass-losing stars and generally attributed to the presence of an expanding mass outflow. These masers have been studied extensively; most recently, astrometry of the H$_2$O masers by Choi et al.\ (2008) was used to derive a precise distance measurement of 1.14$^{+0.11}_{-0.09}$ kpc, a revision of the previous 1.5 kpc distance derived from its assumed membership in the star cluster NGC 2362 (Lada \& Reid 1978). Finally, VY CMa has a reported mass loss rate that is much higher than other RSGs in the Milky Way, including its fellow dust-enshrouded supergiants VX Sgr, S Per, and NML Cyg (Schuster et al.\ 2006).

The spectrum of VY CMa has been monitored carefully for nearly 50 years (Wallerstein 1958). It has appeared to stay at a quite constant spectral type of M3-M5 I (Wallerstein \& Gonzalez 2001), but also displays variable low-exitation emission features that are variable on a timescale of months (Wallerstein 1958, Herbig 1970b) as well as TiO band heads in weak emission (Hyland et al.\ 1969, Wallerstein \& Gonzalez 2001). VY CMa has previously been assigned a very cool $T_{\rm eff}$ of 2800-3000 K (Le Sidaner \& Le Bertre 1996, Smith et al. 2001, Monnier et al.\ 2004, Humphreys et al.\ 2005), which has in turn implied a very high $M_{\rm bol}$ of $-8.5$ to $-9.5$ and placed VY CMa well to the right of the Hayashi limit on the H-R diagram, in the ``forbidden" region where stars are not expected to be in hydrostatic equilibrium. These parameters also implied that VY CMa had an extraordinarily large radius of 1800-3000R$_{\odot}$.

Massey et al.\ (2006b) recently applied the MARCS stellar atmosphere models to spectrophotometry of VY CMa and derived a $T_{\rm eff}$ of 3650 $\pm$ 25 K, much warmer than previous estimates. This implied a much lower $M_{\rm bol} \sim -7.0$ and smaller radius of $\sim600R_{\odot}$ as a result, bringing VY CMa into much better agreement with the predictions of the evolutionary tracks. Humphreys et al.\ (2007) argued that this luminosity was far too low, citing a much higher luminosity of $M_{\rm bol} \sim -9.5$ derived from an integration of VY CMa's spectral energy distribution. This higher luminosity should include the contribution of the central star as well as re-radiated thermal emission from the circumstellar dust; however, Levesque et al.\ (2009a) note that this argument rests on the assumption that VY CMa's dust envelope is spherically symmetric. Finally, Choi et al.\ (2008) estimate $M_{\rm bol} \sim -9.0$ based on their newly revised distance to the star.

Figure 4a summarizes VY CMa's changing position on the Milky Way H-R diagram over the past decade, and compares it to the larger sample of Galactic RSGs from Levesque et al.\ (2005). It is clear that further investigations of this star's unusual physical properties are required in order to gain an accurate understanding of the central star and the surrounding nebula.

\subsubsection{WOH G64}
WOH G64 is an unusual RSG in the LMC that was originally discovered by Westerlund et al.\ (1981). Like VY CMa, it is quite bright in the infrared and surrounded by a dusty nebula (Elias et al.\ 1986, Roche et al.\ 1993, Ohnaka et al.\ 2008). As a source of OH, SiO, and H$_2$O masers (Wood et al.\ 1986, van Loon et al.\ 1996, 1998b, 2001), it has been found to have a considerable mass outflow and two expanding dust shells (Marshall et al.\ 2004). While a binary scenario cannot be ruled out for WOH G64, so far no evidence has supported this possibility (Levesque et al.\ 2009a). The spectrum of WOH G64 is dominated by strong TiO absorption features, leading to its designation as a very cool RSG with a spectral type as late as M7-8 (Elias et al.\ 1986, van Loon et al.\ 2005). The cold temperatures associated with these spectral types ($\sim$3000 K) placed WOH G64 in the same ``forbidden" region on the right-hand side of the H-R diagram that VY CMa previously occupied, and corresponded to an extraordinarily large radius ($\sim$2500 R$_{\odot}$).

Recent studies of WOH G64's dust envelope and physical properties have, like VY CMa, moved it into better agreement with the predictions of the evolutionary tracks. Ohnaka et al.\ (2008) computed detailed models of the star's circumstellar environment based on N-band observations, proposing a geometry that described this envelope as an asymmetric dusty torus being viewed almost head-on. This model led to $M_{\rm bol} = -8.9$, a factor of 2 lower than the luminosities proposed in Elias et al.\ (1986) and van Loon et al.\ (2005). Levesque et al.\ (2009a) used the MARCS stellar atmosphere models to determine $T_{\rm eff} =$ 3400 $\pm$ 25 K and $M_{\rm bol} = -9.4$. This $M_{\rm bol}$ was initially at odds with the luminosity derived by Ohnaka et al.\ (2008). However, the geometrical model proposed by Ohnaka et al.\ (2008) predicts that the dusty torus contributes $\sim$0.5 mag in the $K$ band, the regime that Levesque et al.\ (2009a) used for deriving $M_{\rm bol}$ (since RSGs are found to remain quite constant at $K$; Josselin et al.\ 2000). Taking this into account, Levesque et al.\ (2009a) arrived at a final $M_{\rm bol} = -8.9$, in excellent agreement with that of Ohnaka et al.\ (2008). While these parameters bring WOH G64 into better agreement with the predictions of the LMC-metallicity evolutionary tracks and correspond to a much smaller radius of $\sim$1500 R$_{\odot}$, it is still the coldest RSG in the LMC and one of the largest RSGs ever observed.

One unique feature of WOH G64 lies it its surprising nebular emission line spectrum. In their spectrum, Elias et al.\ (1986) reported detections of [O I] $\lambda$6300, H$\alpha$, [N II] $\lambda\lambda$6548,6584, and potential [S II]$\lambda\lambda$6717, 6731 emission features in their spectrum. Levesque et al.\ (2009a) detect these same emission lines along with H$\beta$, [N I]$\lambda\lambda$5198,5200, and [O III]$\lambda$5007 emission features, as well as weak TiO band heads in emission similar to those observed in VY CMa. The excitation mechanism for these nebular emission lines currently remains unexplained. Line flux ratios and nitrogen abundances both support the possibility of collisional excitation via shock heating. It is also possible that these lines could be excited through ionization by a hot companion, as is seen in interacting RSG spectroscopic binaries such as VV Cep; however, observations of the nebula emission velocities do not support a binary hypothesis. More detailed observations of this unusual RSG, including further spectral analyses and observations in the blue and near-UV regimes, are necessary to fully understand its extreme physical properties and unusual spectral signature. Figure 4b illustrates the changes in WOH G64's position on the LMC H-R diagram over the past several years, and compares it to the LMC RSGs from Levesque et al.\ (2006).

\begin{figure}
\includegraphics[width=8cm]{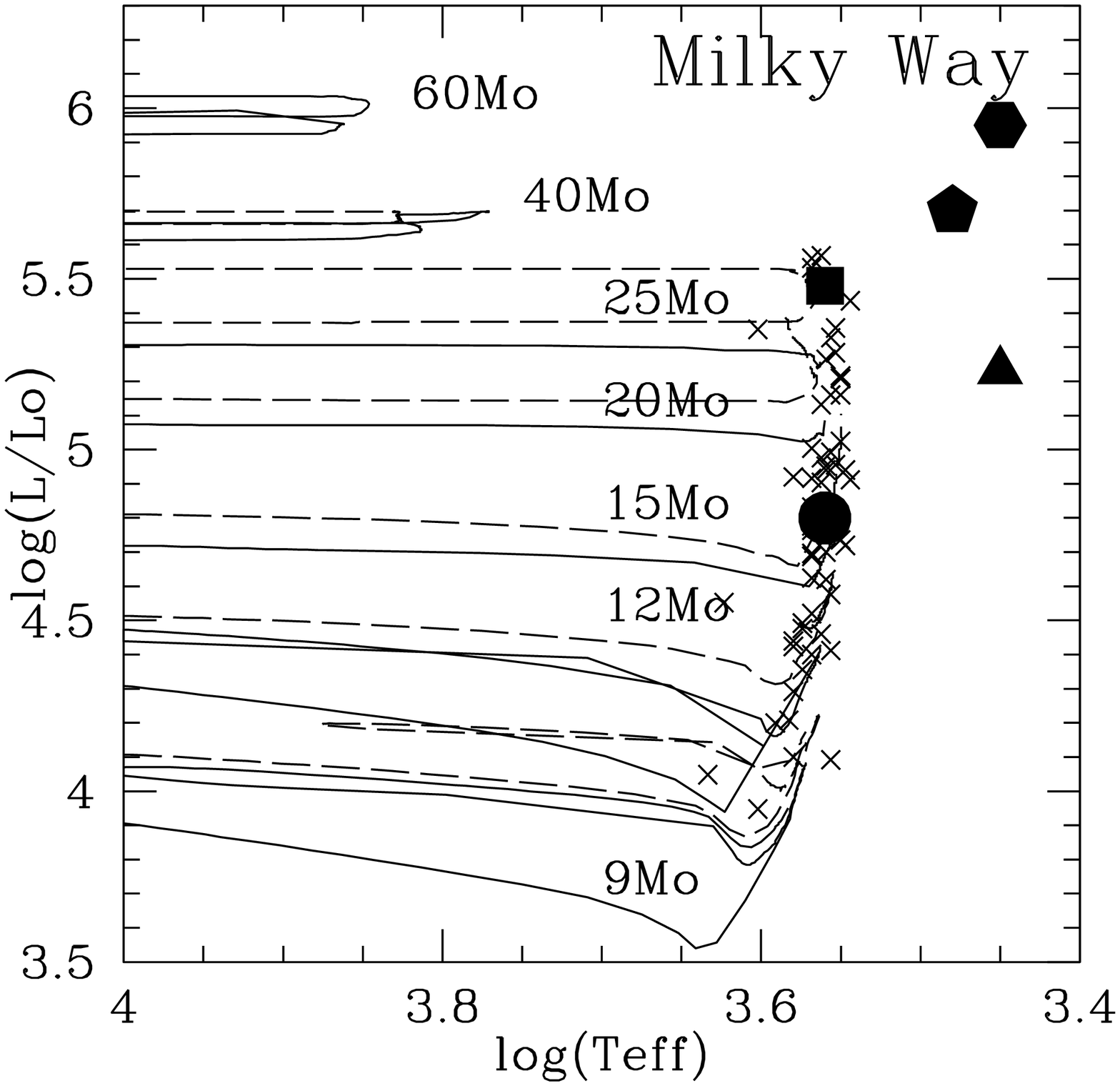}
\includegraphics[width=8cm]{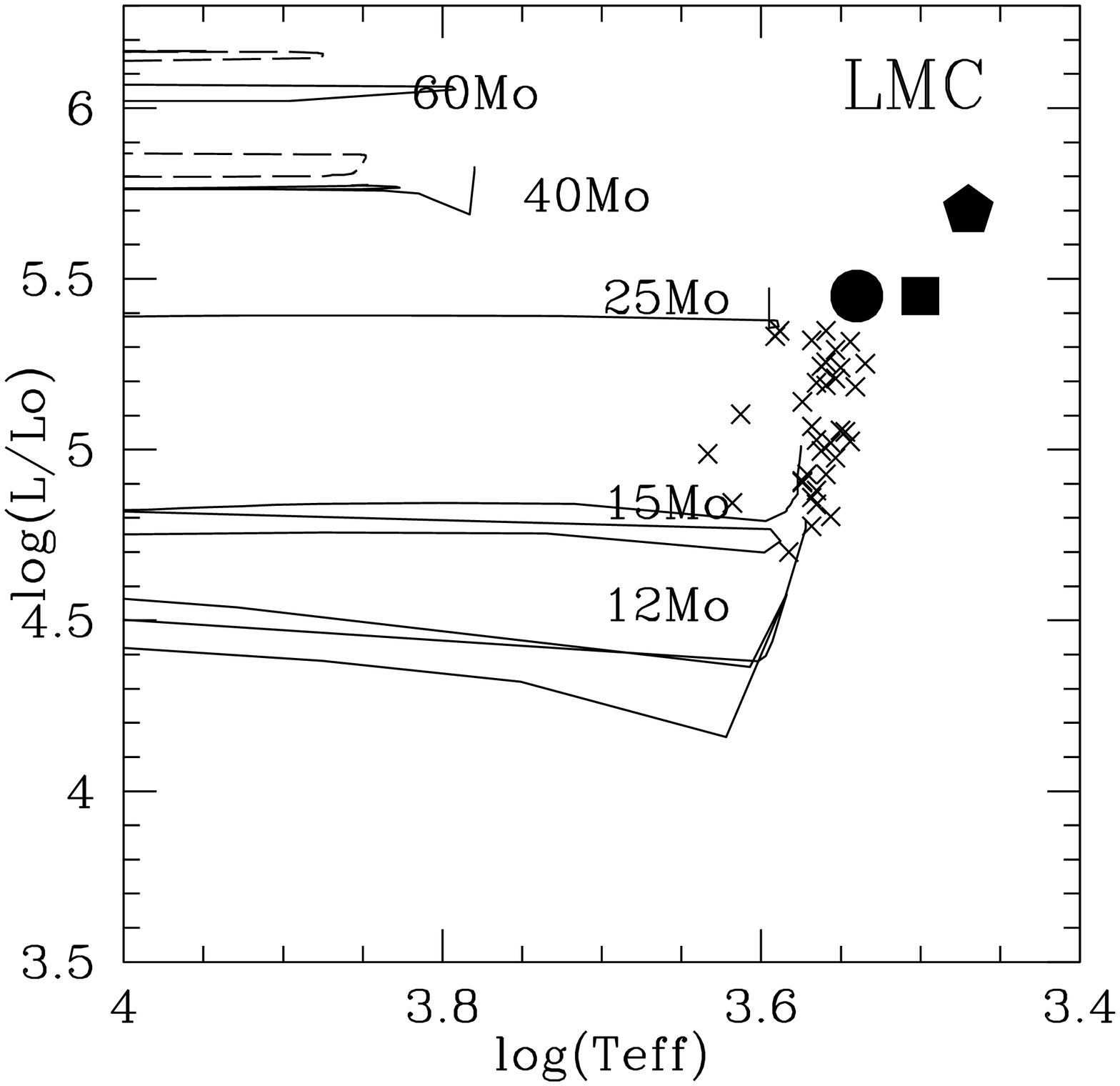}
\caption{Placing VY CMa and WOH G64 on the H-R diagram, adapted from Massey et al.\ (2006b) and Levesque et al.\ (2009a). Figure 4a (left): Position of VY CMa on the Milky Way H-R diagram as determined by several recent studies: Le Sidaner \& Le Bertre (1996; hexagon), Smith et al.\ (2001; pentagon), Monnier et al.\ (1999; triangle), Massey et al.\ (2006b; circle), and Choi et al.\ (2008; square). For comparison, the sample of Galactic RSGs from Levesque et al.\ (2005) are shown as crosses. The evolutionary tracks are taken from Meynet \& Maeder (2003). Figure 4b (right): Position of WOH G64 on the LMC H-R diagram as determined by recent studies: van Loon et al.\ (2005; pentagon), Ohnaka et al.\ (2008; square), and Levesque et al.\ (2009a; circle). For comparison, the sample LMC RSGs from Levesque et al.\ (2006) are shown as crosses. The evolutionary tracks are taken from Meynet \& Maeder (2005) and include both non-rotating tracks (solid lines), and tracks that assume an initial rotation velocity of 300 km s$^{-1}$ (dashed lines).}
\end{figure}

\subsubsection{VX Sagittarii, S Persei, and NML Cygni}
Aside from VY CMa, the most well-studied dust-enshrouded supergiants in the Milky Way are, VX Sagittarii, S Persei, and NML Cygni. Photometric observations of these stars are quite extensive, with data spanning $\sim$50-100 years. However, observations of these stars' optical and near-IR spectra have not been updated since initial determinations of their spectral types. These were based on optical and near-IR spectrograms (e.g. Bidelman 1947, Wing et al.\ 1967, Low et al.\ 1970, Herbig \& Zappala 1970, Humphreys \& Lockwood 1972, Humphreys 1974), narrow-band eight-color spectra (White \& Wing 1978), and spectroscopy (Lockwood \& Wing 1982). These methods utilize a wide variety of spectral features and techniques to determine spectral type, and extrapolations of the stars' physical properties are based on calibrations for {\it giant} stars (Ridgway et al.\ 1980, Lockwood \& Wing 1982) and did not account for what we now know to be potentially critical effects of the asymmetric dust reflection nebulae on determinations of luminosity for the central RSG. In recent years our understanding of Galactic RSGs and their physical properties has seen several improvements and updates through observational studies (Levesque et al.\ 2005), more accurate stellar evolutionary tracks (Schaller et al.\ 1992, Meynet et al.\ 1994), and the new generation of the MARCS stellar atmosphere models (Plez 2003, Gustafsson et al.\ 2003, Gustafsson et al.\ 2008). S Per, VX Sgr, and NML Cyg would benefit enormously from new spectrophotometric observations that could be used to reevaluate these stars' physical properties.

However, the recent studies of VY CMa and the LMC star WOH G64 have highlighted that such analyses are strongly dependent on accurate and detailed observations of the thick dust envelopes surrounding these stars. Fortunately, in recent years there {\it have} been several excellent imaging studies that closely probed the geometry of the dusty nebulae surrounding S Per, VX Sgr, and NML Cyg, conducted by Monnier et al.\ (2004) and Schuster et al.\ (2006, 2009). As demonstrated by VY CMa and WOH G64, these observations will be vital in accommodating the effects of the dust nebula and determining accurate physical properties for the central stars. Combining these new circumstellar geometries with new spectral types and effective temperatures for these stars will be vital for properly determining these stars' luminosities, placing them accurately on the H-R diagram, and scrutinizing how they fit into our current picture of RSG evolution and the Galactic RSG population.

\subsection{Variable Red Supergiants}
As discussed in Section 3.2, the H-R diagram shows a shift of the Hayashi limit to warmer temperatures at lower metallicities. This rightmost reach of the evolutionary tracks should impose a hard limit on how cold RSGs can be, and hence how late their spectral types can be, in a particular environment. The expected result is a lack of cold, late-type stars in lower-metallicity galaxies such as the Magellanic Clouds.

Despite this, Levesque et al.\ (2007) found that observations of Magellanic Cloud RSGs in November 2004 revealed several RSGs whose spectral subtypes appeared unusually late with respect to the average type for their host. Additional observations in December 2005 revealed large discrepancies in the spectral subtypes assigned to several of the stars over this $\sim$1 year timescale. While determining spectral types has always involved a small degree of subjectivity, there is no comparable disagreement seen between, for examples, the Massey \& Olsen (2003) spectral types and Levesque et al.\ (2006) spectral types for the same sample of Magellanic Cloud RSGs. In fact, spectral variability of a type or more is unheard of in the general RSG population (multiple detailed spectrophotometric observations are not currently available to confirm potential variability in the OH/IR supergiant population). These differences in spectral type were verified by directly comparing the 2004 and 2005 spectra of these RSGs; substantial changes are apparent in the strengths of the broad TiO absorption features in the RSG spectra.

The most dramatic example of this variation is the RSG HV 11423 in the SMC, described by Massey et al.\ (2007b). Originally observed in November 2004, the star was assigned a spectral type of K0-1 I and a $T_{\rm eff}$ of 4300 K. However, observations from December 2005 revealed that the star's spectrum had changed significantly, with a much later spectral type of M4 I and a much cooler $T_{\rm eff}$ of 3500 K (see Figure 5). A third spectrum in the blue was observed in September 2006. This showed that the spectrum had changed yet again and now appeared to be in excellent agreement with the K0-1 I spectrum from 2004. The December 2005 spectral type of M4 I is by far the latest type assigned to an RSG in the SMC, and well later than the average spectral type of K5-M0 I (Massey \& Olsen 2003, Levesque et al.\ 2006). In addition to these extreme and rapid variations in spectral type, HV 11423 displays abnormally high variability in $V$, well in excess of the $\sim$1 mag variations typical of RSGs (Josselin et al.\ 2000, Levesque et al.\ 2007), and also shows considerable variations in $M_{\rm bol}$ and $A_V$, appearing brightest, dustier, and more luminous in its warm early-type state.

\begin{figure}
\includegraphics[width=15cm]{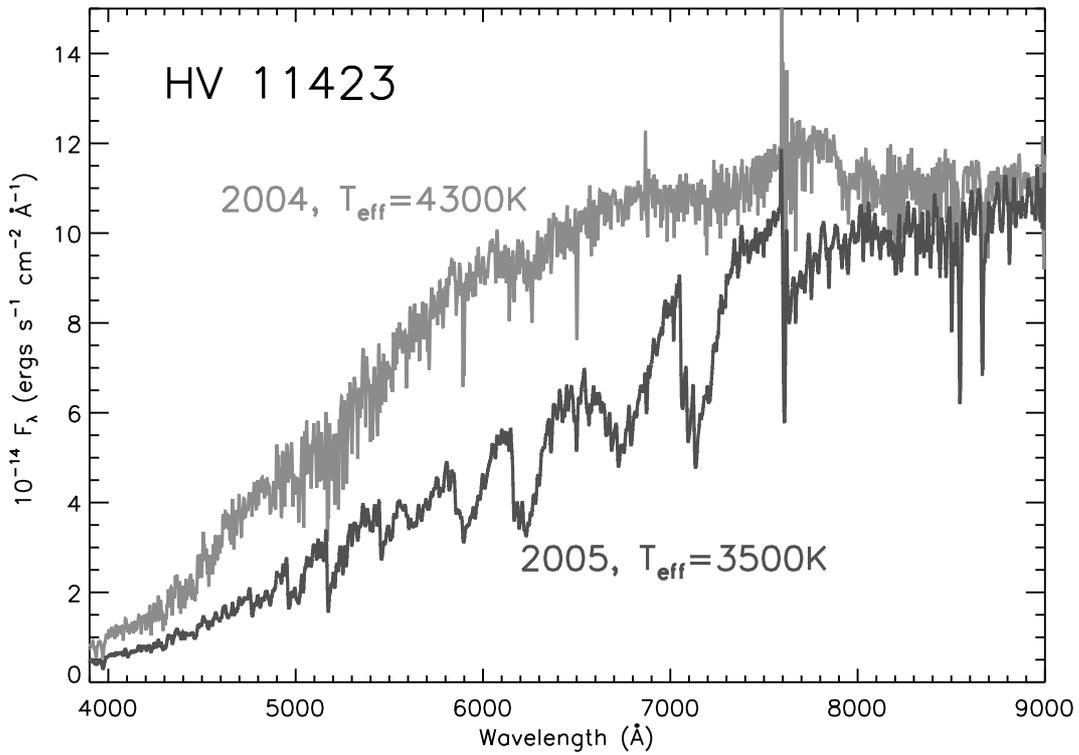}
\caption{The changing spectrum of HV 11423, adapted from Massey et al.\ 2007b. The 2004 spectrum of HV 11423 (light gray) is shown to have a cool $T_{\rm eff}$ of only 4300 K and a spectral type of K0-1 I. By comparison the 2005 spectrum (dark gray) has a much cooler $T_{\rm eff}$ of 3500 K, corresponding to a spectral type of M4 I, and shows a drastically different spectrum, with strong TiO band. No adjustments in flux have been made for these observations, showing that HV 11423 was significantly brighter in 2004. The strong feature at 7600\AA\ is the telluric A band.}
\end{figure}

Intriguingly, Levesque et al.\ (2007) found that three more RSGs in the Magellanic Clouds - two in the SMC and one in the LMC - fit into precisely the same behavioral template. All of these stars, the Levesque-Massey (L-M) variables, exhibit unusual variability in their optical spectrum, $V$ magnitude, $M_{\rm bol}$, and $A_V$ on the timescales of months, and the stars are brighter, dustier, and more luminous when they display their earliest spectral types. These variations in extinction are characteristic of the effects of circumstellar dust, and could be connected with sporadic dust production from these stars, a  phenomenon that has been previously described by Danchi et al.\ 1994 based on studies of circumstellar dust envelopes around RSGs. Finally, while effective temperatures of RSGs determined using the MARCS models are consistent with the evolutionary tracks in the Milky Way,
Magellanic Clouds, and M31 (including the dust-enshrouded RSGs VY CMa and WOH G64), no such agreement is seen for the L-M variables. In their coolest states, the $T_{\rm eff}$ determined for each of these stars from the MARCS models still remains at odds with the predictions of the evolutionary tracks, lying to the right of the Hayashi limit on the H-R diagram.

Variability of this magnitude and on this surprisingly short timescale has never been previously confirmed in RSGs, and the current belief is that these unusual properties are indicative of an unstable, and likely short-lived, evolutionary phase not previously associated with massive stars. This unusual behavior appears to be due in part to the warmer limits of hydrostatic equilibrium imposed by these stars' lower-metallicity environments. However, it is still not clear whether these variations represent physical changes in the star's atmosphere or apparent changes imposed by sporadic dust production episodes and consequential effects on the stars' optical depths and apparent spectra (Meynet, private communication). Rigorous and continuous follow-up observations of these stars' optical spectra, as well as infrared observations of their dust properties, are necessary in order to form a detailed picture of this variable behavior and its ultimate origin.

\section{Red Supergiants: Current Questions and Future Work}

\subsection{RSGs and the H-R Diagram}
RSGs in the Milky Way, the Magellanic Clouds, and M31 now show excellent agreement with the predictions of the evolutionary tracks (Levesque et al.\ 2005, 2006, Massey et al.\ 2009). From these detailed examinations of RSG physical properties across a range of metallicities, we are beginning to understand the effects that metallicity can have on RSG physical properties, populations, and lifetimes. However, there are still many questions that remain unanswered about RSGs and the effects of metallicity:
\begin{itemize}
\item{Do very low-metallicity evolutionary tracks accurate produce the physical properties of RSGs?}
\item{Does the progression of observed B/R and RSG/WR ratios continue to even lower metallicities?}
\item{How does metallicity affect the range of initial masses for stars that will eventually pass through an RSG stage and, in some cases, evolve back towards the blue?}
\end{itemize}

Many of these questions can be investigated through observations of RSG populations in the Local Group and other nearby galaxies. The Local Group Survey presented in Massey et al.\ (2007a) has identified RSG populations in seven nearby dwarf star-forming galaxies, and spectroscopic follow-up is now required in order to confirm these stars as RSGs and members of their proposed host galaxies. The sample of RSGs in the Magellanic Clouds has already been studied in detail, though additional observations would of course be beneficial. Extending these same studies to the low-metallicity dwarf galaxies NGC 6822, WLM, and Sextans A and B would be extremely valuable in answering many of the existing questions about the role that metallicity plays in RSG populations. Studies of the massive star population in IC 10 would be particularly intriguing; Massey et al.\ (2007a) find a surprising dearth of RSGs relative to the blue supergiant and W-R population in these galaxies, placing IC 10 at odds with the observed increase of the RSG/W-R ratio at lower metallicities (Massey 2002, Massey \& Holmes 2002; see also Section 3.2.2). Massey et al.\ (2007a) speculate that this is due to a current ongoing starburst phase in IC 10 (Massey \& Armandroff 1995). This explanation is quite intriguing, as it could have important implications for proper modeling of massive star populations in galaxies with current starburst activity or burst-like star formation histories.

Physical properties for these low-metallicity Local Group RSGs would extend our understanding of massive stellar evolution to unprecedentedly low metallicities, and provide a means of testing the accuracy of stellar evolutionary tracks at these metallicities for the first time. In conjunction with spectroscopic surveys of early-type supergiants in these Local Group galaxies (such as, for example, studies of O, B, and A supergiants in the dwarf irregular galaxy WLM by Bresolin et al.\ 2006 and Urbaneja et al.\ 2008), we could evaluate the B/R and RSG/WR ratios at these metallicities. Finally, a larger sample size of RSGs throughout the Local Group would permit a more rigorous determination of $L_{\rm max}$ across these galaxies, perhaps leading to a better understanding of the purportedly ``missing" 25-40M$_{\odot}$ RSGs and the surprising result that $L_{\rm max}$ does not appear to be dependent on metallicity (see Section 3.2.2).

\subsection{Dust Production in RSGs}
Several recent studies of RSGs have revealed the importance and complexity of their mass loss and dust production. The presence of circumstellar dust is common in the general RSG population, and studies of this dust have revealed evidence of episodic mass loss events (Danchi et al.\ 1994). RSG mass loss rates appear to be dependent on luminosity and are expected to be high enough to dominate dust production in young galaxies at large look-back times, where an underlying dust-producing population of AGB stars will not yet have had time to form (Massey et al.\ 2005). Finally, there is preliminary evidence to suggest that the dust produced by RSGs does not appear to follow the standard $R_V = 3.1$ Cardelli et al.\ (1989) reddening law of the diffuse ISM (Bennett et al.\ 2009). These recent advances had highlighted the importance of several questions that still remain to be answered about RSGs and their dust production:
\begin{itemize}
\item{How do RSG mass loss rates change with metallicity?}
\item{What are the various mechanisms that drive RSG mass loss and contribute to its apparent episodic nature?}
\item{What reddening law does the dust produced by RSGs follow?}
\end{itemize}

Answering these questions will require detailed observations of RSG samples at a variety of wavelengths, including the IR (to more carefully study the circumstellar dust shells around these stars and their mass loss rates), and the UV (to facilitate multi-wavelength studies of RSG spectra and determine the reddening law followed by the circumstellar dust).

It is critical to fully understand the mass loss processes in RSGs. Evolutionary channels for massive stars are extremely dependent on mass loss rates and mechanisms; evaluating the proposed progenitor models for supernovae and gamma-ray bursts are extremely dependent on a complete understanding of the ways in massive stars shed mass. It is also extremely important to quantify the reddening law followed by RSG-produced dust. Since these stars are expected to dominate dust production at large lookback times, evidence of a non-standard reddening law could have far-reaching implications for interpretations of reddening effects in the high-redshift universe.

\subsection{Unusual RSGs}
We have seen in this review that dust-enshrouded supergiants are an intriguing subset of extreme RSGs. These stars are characterized by their thick dust envelopes, which complicate accurate determinations of the central stars' physical properties. Another sample of unusual RSGs, the L-M variables, have recently been observed in the Magellanic Clouds. These stars show dramatic variability in their spectra on the timescale of months. This unusual behavior coincides with what appear to be sporadic mass loss episodes, and seems to be at least partially dependent on the metallicity-dependent limits of hydrostatic equilibrium, suggesting that these variables are more common at lower metallicities. Future studies of these unusual RSGs should aim to answer several outstanding questions:
\begin{itemize}
\item{Where do dust-enshrouded supergiants fit into the H-R diagram and our current understanding of RSG evolution?}
\item{What is the physical explanation behind the observed variability of L-M variables?}
\item{Is L-M variability in RSGs dependent on metallicity?}
\end{itemize}

Properly interpreting the effects that circumstellar environments can have on the determination of dust-enshrouded RSG physical properties is largely dependent on an accurate understanding of the dust shell geometry. Accurate physical properties also require spectrophotometric observations that can be used to determine $T_{\rm eff}$ and $M_{\rm bol}$. Recent studies of VY CMa and WOH G64 have illustrated that such analyses are possible; however, the Galactic RSGs S Per, VX Sgr, and NML Cyg still remain to be examined. Fortunately, excellent new observations by Schuster et al.\ (2006, 2009) have recently been carried out that probe the dust morphologies around these stars in detail. Following these analyses, the time is right for new spectrophotometric observations of these stars. With these data, we would be able to infer physical properties of the central stars while accommodating for the effects of the dust nebulae. Dust-enshrouded supergiants could present critical examples of extreme mass loss mechanisms and dust production, so including them in our current understanding of RSG evolution is particularly important.

A proper investigation of L-M variables will require a regular monitoring program for the existing variables that have been detected in the LMC and SMC. This will permit a more complete description of their variability, including any potential evidence for periodicity or large-scale changes in the stars, and provide the information that is necessary to determine a physical explanation for this unusual behavior. Increasing the current sample size of L-M variables would also be beneficial to these studies and help to answer the question of metallicity's role in these stars' evolution and behavior. Such a search could begin with photometric observations, as some of these stars (such as HV 11423), show clear changes in their color indices as a result of extreme spectral variability. However, unambiguous detections of other RSGs displaying L-M variability will require multiple spectroscopic observations of RSGs in other low-metallicity Local Group galaxies, such as NGC 6822 and WLM\@. Detections of RSGs with unusually late spectral types, as well as further observations of short-term spectral variability, will help to further our understanding of the environments and physical phenomena governing the behavior of these unusual stars.

\subsection{RSGs and the Big Picture}
RSGs represent an important phase of massive stellar evolution. In many cases, the RSG phase marks the terminal evolutionary state of a massive star, eventually exploding as a Type II supernova. In other scenarios, RSGs are an important intermediate phase in the life of a massive star, identified primarily as a period of high mass loss before the star evolves leftwards across the H-R diagram and becomes a yellow supergiant or W-R star. The importance of RSGs in both of these scenarios is just now coming to light, and bringing with it a host of questions that will guide studies of RSGs in the coming years:
\begin{itemize}
\item{What mass range of RSGs explode as SNe? What mass range of RSGs evolve back into yellow supergiants or W-R stars? How do these mass ranges change with metallicity?}
\item{How can we identify yellow supergiants and Wolf-Rayet stars that have undergone a previous RSG phase?}
\item{What role does the RSG phase play in the properties of SNe produced by massive stars, both from terminal RSGs or massive stars that have undergone a previous RSG phase?}
\end{itemize}

We currently have only very limited means of drawing conclusions about RSG masses in different galaxies and at different metallicities. For example, determinations of $L_{\rm max}$ are inconclusive since luminosity and mass become degenerate at the right-hand end of the H-R diagram. We also cannot state conclusively whether changes in the RSG/W-R ratio with metallicity can be attributed to changing lifetimes of these evolutionary phases or changes in the mass ranges that produce RSGs and W-R stars. Answering these questions will prove difficult until the metallicity-dependent physical properties of RSGs have been studied in detail across the Local Group. There does appear to be some evidence illustrating how ``post"-RSG massive stars can be identified. Oudmaijer et al.\ (2008) present observations of two yellow supergiants with substantial circumstellar material, evidence of a previous high-mass loss phase. However, additional means of identifying yellow supergiants or W-R stars that have undergone a previous RSG phase still require investigation. Finally, understanding the mass ranges and evolutionary effects of the RSG phase are both critical when considering the final fate of these massive stars as core-collapse SNe. It would be quite beneficial to understand the effect that RSG mass loss processes have on the eventual production of Type II or Type Ibc SNe.

Our understanding of RSGs, their physical properties, and their role in massive stellar evolution has grown dramatically in the past 10-15 years. At the same time, the complexities of these stars have become increasingly numerous, highlighted in the observed effects of metallicity on their physical properties and lifetimes, their high mass loss rates and circumstellar dust properties, and the enigmatic dust-enshrouded supergiants and L-M variables that complicate our understanding of mass loss mechanisms and stellar evolutionary theory. Our new understanding of RSGs, their physical properties, and their role in stellar evolution also brings with it a number of new and intriguing questions that are ripe for investigation in the years to come.

Ongoing collaboration and correspondence with Phil Massey was invaluable during the preparation of this manuscript. In addition, this review has benefited from collaborations and conversations with Phil Bennett, Geoff Clayton, Peter Conti, Eric Josselin, Andre Maeder, Georges Meynet, Knut Olsen, Bertrand Plez, David Silva, and Brian Skiff. This work has been supported in part through a Ford Foundation Predoctoral Fellowship.

\end{document}